\documentclass{amsart}
\usepackage{amsmath}
\usepackage{amssymb}
\usepackage{amsxtra}
\begin{document}
\newtheorem{theorem}{Theorem}
\newtheorem{axiom}{Axiom}
\newtheorem{corollary}{Corollary}
\author{Ilja Schmelzer}
\email{ilja.schmelzer@gmail.com}%
\urladdr{ilja-schmelzer.de}
\thanks{Electromagnetic Phenomena, vol.3, nr.1 (9), 2003}

\title{A Fermion Doublet With Chiral Gauge Interaction On A Lattice} 
\sloppypar\sloppy

\maketitle
\begin{abstract}
We present a new staggered discretization of the Dirac operator.
Doubling gives only a doublet of Dirac fermions which we propose to
interpret as a physical (lepton or quark) doublet.  If coupled with
gauge fields, an $(1+\gamma^5)$ chiral interaction appears in a
natural way.  We define a generalization for curved background which
does not require tetrad variables.  The approach suggests a natural
explanation for the three fermion families.
\end{abstract}

\tableofcontents

\section{Introduction}

In this paper we present a new discretization of the Dirac equation.
In comparison with staggered fermions \cite{Kogut} it creates not four
but only two flavours of Dirac fermions.  This has been reached by
placing not only different spin components, but also their real and
imaginary parts into different nodes.  These sixteen degrees of freedom
(two fermions) can be understood, in some sense, as the result of
``doubling'' of a real scalar step of freedom $\varphi(n)$ on the
lattice.

Moreover, these two fermions live on different sub-lattices ($\psi_o$
on ``odd'' nodes, $\psi_e$ on ``even'' nodes), thus, we obtain also a
single fermion (eight degrees of freedom) on the lattice by omitting
half of the lattice nodes.  But, if the neutrino is a Dirac particle,
all fermions appear in doublets of Dirac particles.  In this context,
the appearance of a fermion doublet in this discretization may be not
a bug but a feature which allows to explain why Dirac particles appear
in such doublets.

In our approach the complex structure is an operator among others.
Moreover, there is no natural complex structure, but, instead, a
quaternionic structure.  The choice of a complex structure corresponds
to a choice of a direction.  To avoid unnecessary symmetry breaking we
propose to consider a vector field instead of a scalar field on the
lattice.  In this case, each component has its own natural preferred
direction and therefore a natural complex structure.  This gives in a
natural way three generations of our doublet.

Once a complex structure is defined, we can define the operator
$\gamma^5$ on the grid. It exactly anticommutes with $D$.  On the
other hand, $(\gamma^5)^2\ne 1$ but a shift by $2a$ in the preferred
direction.  The subspaces $\gamma^5=\pm 1$ have an uncertain
definition but a clear physical interpretation: they define the
subspaces $\psi_o\approx\pm\psi_e$.  An $1+\gamma^5$ type chiral
interaction with a gauge field appears in a straightforward way from
the typical lattice gauge term $A_{n,n+a_i}(\phi(n)+\phi(n+a_i))$: If
$\gamma^5\approx 1$ it becomes $\approx A_i \phi(n)$, but if
$\gamma^5\approx -1$ it becomes almost zero.

The discretization may be generalized to a curved metric background.
The fermion doublet may be, in this context, described without any
tetrad variables by the sixteen-dimensional exterior bundle of
four-dimensional differential forms $\Omega=\sum_{k=0}^n\Omega^k$.  On
this space exists a natural ``Dirac operator'' (in the sense of a
square root of the harmonic operator) as well as a natural
discretization for a general mesh.

In appendix \ref{ether} we consider the connection of this concept
with a generalization of Lorentz ether theory to gravity.  In this
context, the proposal to explain fermion families as natural degrees of
freedom for ether crystal dislocations becomes really meaningful.
Moreover, we define a natural $SO(3)$ gauge field which measures mesh
irregularity which we propose as a candidate for the $SU(2)$ part of
electroweak theory.

There is a large number of proposals for chiral gauge theories on the
lattice.  They are mainly focussed around the Ginsparg-Wilson relation
\cite{Ginsparg} or its generalizations \cite{Kerler}, especially
domain wall fermions \cite{Shamir}, Neuberger's overlap operator
\cite{Neuberger}, proposals by Fujikawa \cite{Fujikawa} and Chiu
\cite{Chiu}.  Because in our approach $(\gamma^5_{mesh})^2\neq 1$,
there is no straightforward connection between our approach and these
other approaches.

\section{A Real Representation of the Dirac Algebra}

We forget -- for some time -- about the complex structure.  Instead of
the usual representations with four complex fields, we use an
eight-dimensional real representation of the operators $\gamma^\mu$
defined here by their linear combination with $\partial_\mu$:

\begin{equation}
\label{spinor}
\gamma^0\partial_0-\gamma^i\partial_i =_{def}
\left(
\begin{array}{rrrrrrrr}
 \partial_0& \partial_1& \partial_2&           & \partial_3&&&\\
-\partial_1&-\partial_0&           & \partial_2&& \partial_3&&\\
-\partial_2&           &-\partial_0&-\partial_1&&& \partial_3&\\
           &-\partial_2& \partial_1& \partial_0&&&& \partial_3\\           
-\partial_3&&&&-\partial_0&-\partial_1&-\partial_2&           \\
&-\partial_3&&& \partial_1& \partial_0&           &-\partial_2\\
&&-\partial_3&& \partial_2&           & \partial_0& \partial_1\\
&&&-\partial_3&           & \partial_2&-\partial_1&-\partial_0          
\end{array}
\right)
\end{equation}

In the context of this representation, it seems also natural to define
(by their linear combination with scalar parameters $m_i$) the
following operators $\beta^i$:

\begin{equation}
\label{mass_matrix}
m_i\beta^i =_{def}
\left(
\begin{array}{rrrrrrrr}
    & m_1& m_2&           & m_3&&&\\
 m_1&    &           & m_2&& m_3&&\\
 m_2&           &    &-m_1&&& m_3&\\
           & m_2&-m_1&    &&&& m_3\\           
 m_3&&&&    &-m_1&-m_2&           \\
& m_3&&&-m_1&    &           &-m_2\\
&& m_3&&-m_2&           &    & m_1\\
&&& m_3&           &-m_2& m_1&              
\end{array}
\right)
\end{equation}

The following operator equation holds:

\begin{equation}
(\gamma^0\partial_0-\gamma^i\partial_i + m_i\beta^i)^2 
 = -\square + \delta^{ij} m_i m_j
\end{equation}

This can be easily seen -- this operator iterates three times, in each
coordinate direction, the same trick:\footnote{This observation also
suggests how to iterate this construction to arbitrary dimension.}

\begin{equation}
\left(
\begin{array}{cc}
A&(m_i+\partial_i)I\\(m_i-\partial_i)I&-A
\end{array}
\right)^2
=
(A^2+(m_i+\partial_i)(m_i-\partial_i)I)\left(
\begin{array}{cc}
I&0\\0&I
\end{array}
\right)
\end{equation}

As follows immediately, the $\gamma^\mu$ define a representation of
the Dirac matrices, and the matrices $\beta^i$ fulfil the following
anticommutation relations:

\begin{equation}
\beta^i\beta^j+\beta^j\beta^i=\delta^{ij}
\end{equation}

and anticommute with all $\gamma^\mu$:

\begin{equation}
\beta^i\gamma^\mu+\gamma^\mu\beta^i=0
\end{equation}

It is also easy to see (and to generalize to arbitrary dimension) that

\begin{equation}
\gamma^0(\gamma^1\beta^1)(\gamma^2\beta^2)(\gamma^3\beta^3)=1.
\end{equation}

The ``classical'' operator $\gamma^5=
-i\gamma^0\gamma^1\gamma^2\gamma^3$ depends on the complex structure,
which is not yet defined.  The natural replacement which does not
depend on it -- the expression $\gamma^0\gamma^1\gamma^2\gamma^3$ --
we denote with $\iota$:

\begin{equation}
\iota=_{def}\gamma^0\gamma^1\gamma^2\gamma^3=\beta^1\beta^2\beta^3 \hspace{1cm}
\iota\gamma^\mu+\gamma^\mu\iota=0\hspace{1cm}
(\iota)^2 = -1
\end{equation}

\section{Discretization of the Dirac Equation}

This representation is appropriate for a discretization of the Dirac
equation on a regular hyper-cubic lattice.  It can be obtained in a
quite simple way: We start with a naive central difference
approximation

\begin{equation}
\partial_i\psi(n)\to\frac{1}{2a_i}(\psi(n+a_i)-\psi(n-a_i)).
\end{equation}

This naive discretization leads to the problem of ``fermion
doubling''.  The continuous limit of this set of discrete equations
gives not only the original Dirac equation, but also additional,
highly oscillating components, the ``doublers''.  In classical
computations such doublers may be often ignored, but in quantum
computations, where the number of degrees of freedom is important (Pauli
principle) this is no longer possible.  We obtain in each direction a
factor to, thus, $2^4=16$ doublers.  Fortunately, eight pairs of
doublers decouple in a really simple way: It is sufficient to hold
only one of the eight real components $\psi^a$ per node.  On the
three-dimensional reference cube
$(\varepsilon_1,\varepsilon_2\varepsilon_3), \varepsilon_i\in\{0,1\}$
we obtain the following locations for the eight components:

\begin{equation}
\label{location3D}
\begin{array}{@{\psi}l@{\mbox{ located at }(}c@{),\hspace{1cm} \psi}l@{\mbox{ located at }(}c@{);}}
^0&0,0,0&^4&0,0,1\\
^1&1,0,0&^5&1,0,1\\
^2&0,1,0&^6&0,1,1\\
^3&1,1,0&^7&1,1,1
\end{array}
\end{equation}

What remains are sixteen degrees of freedom (eight degrees of freedom on
two time steps which we need because of our use of central differences
in time) which corresponds to a doublet of Dirac fermions\footnote{
This is a variant of a well-known approach to solve the ``fermion
doubling'' problem -- so-called staggered fermions \cite{Kogut}.  This
approach reduces the doublers only by a factor four.  Because we
ignore the complex structure of the standard representation, we are
free to place ``real'' and ``imaginary'' part of the complex fields
into different nodes. This gives the additional reduction by factor
two.}.  Note that our discretization may be interpreted as a way to
discretize the d'Alembert equation for a single scalar step of freedom
$\varphi(n)$ with central differences, which gives $2^4 = 16$
doublers.

Now, the last doublet decouples too, but in a slightly less trivial
way: We can distinguish ``even'' and ``odd'' nodes on the full
space-time lattice.  The central difference equations on even (odd)
nodes connects only values on odd (even) nodes.  Thus, we obtain two
fermions $\psi_e$ and $\psi_o$ on even resp. odd nodes.  On the
four-dimensional reference cube
$(\varepsilon_0,\varepsilon_1,\varepsilon_2\varepsilon_3),
\varepsilon_i\in\{0,1\}$ we have

\begin{equation}
\begin{array}{@{\psi_e}l@{\mbox{ located at }(}c@{),\hspace{1cm} \psi_e}l@{\mbox{ located at }(}c@{);}}
^0&0,0,0,0&^4&1,0,0,1\\
^1&1,1,0,0&^5&0,1,0,1\\
^2&1,0,1,0&^6&0,0,1,1\\
^3&0,1,1,0&^7&1,1,1,1
\end{array}
\end{equation}

\begin{equation}
\begin{array}{@{\psi_o}l@{\mbox{ located at }(}c@{),\hspace{1cm} \psi_o}l@{\mbox{ located at }(}c@{);}}
^0&1,0,0,0&^4&0,0,0,1\\
^1&0,1,0,0&^5&1,1,0,1\\
^2&0,0,1,0&^6&1,0,1,1\\
^3&1,1,1,0&^7&0,1,1,1
\end{array}
\end{equation}

But, instead of removing one sub-mesh to describe a single Dirac
fermion, we propose to accept above doublers as a way to describe a
physically meaningful flavour doublet.  Remarkably, if the neutrino is
a standard Dirac particle, then all fermions of the standard model
appear in doublets.  The appearance of a fermion doublet in our
approach may be, therefore, not a bug but a feature which allows to
explain the existence of these doublets.

\section{Complex Structures}

To connect the Dirac particle with a gauge field we need a complex
structure -- a multiplication with $i$.  In our real representation
this is an operator. It's basic property is $i^2=-1$.  To be
compatible in the usual way with the Dirac equation, we also need
$[\gamma^\mu,i]=0$.  This does not define the complex structure
uniquely.  We have several interesting candidates for a complex
structure:

\begin{eqnarray}
i &= \beta^1\beta^2 &= \iota\beta^3\\
j &= \beta^2\beta^3 &= \iota\beta^1\\
k &= \beta^3\beta^1 &= \iota\beta^2
\end{eqnarray}

which together define a quaternionic structure:\footnote{The classical
representation $ij=k$ can be obtained using reverse signs for $i,j,k$,
but we prefer this sign convention because it gives
$\gamma^5=\beta^3$.}

\begin{equation}
i j = - j i = -k;\; 
j k = - k j = -i;\; 
k i = - i k = -j;\; 
i^2 = j^2 = k^2 = -1
\end{equation}

For each candidate for a complex structure, we obtain an own operator
$\gamma^5=_{def}-i\iota$.  Especially for $i=\iota\beta^3$ we obtain
$\gamma^5=-\iota\beta^3\iota=\beta^3$.

\subsection{Fermion Families and Lattice Distortions}

Thus, it seems that to fix the complex structure we somehow have to
break spatial symmetry.  But there is a simple way out of this.
Instead of one scalar step of freedom $\varphi(n)$ on the lattice, we
can consider a vector field -- thus, three components $\varphi^i(n)$.
Now, each component has a natural ``preferred direction'' and,
therefore, a natural complex structure.

Moreover, a vector on a lattice is a quite natural step of freedom.
It is, for example, the natural way to describe lattice distortions
with a shift vector field $u^i(n)$.

On the other hand, in the standard model we have three families --
three copies of each fermion.  This suggests to explain on the
kinematic level the three fermion families using the hypothesis that
the fundamental degrees of freedom of the ``theory of everything'' are
three-dimensional vector fields $u^i(n)$.  This idea becomes really
meaningful only in an ether-theoretical context, where relativistic
symmetry is considered as not fundamental but an effective large scale
symmetry, and distortions of an ``ether crystal'' are natural degrees of
freedom.  We consider such an approach in more detail in appendix
\ref{ether}.

For our symmetry problem -- the choice of the complex structure --
other solutions are possible.  We can use them all in expressions like

\begin{equation}
i A^1 + j A^2 + k A^3
\end{equation}

which allows to connect with an $SU(2)$ or $SO(3)$ gauge field.  We
can also try to connect colors with spatial directions to define the
preferred complex structure differently for each colored quark.  Last
not least, it may be really broken (possibly the $U(1)$ part of
electroweak interaction).  These different solutions possibly even do
not contradict each other.  

But it seems important to notice that the choice of the complex
structure is not natural, requires a symmetry breaking of the
quaternionic structure.  This point may be important for the
understanding of symmetry breaking in the standard model.  

\section{Chiral Symmetry on the Lattice}

Let's assume now that one of the complex structures, namely
$i=\iota\beta^3$, has been chosen.  To understand chiral symmetry we
have to define the operator $\gamma^5=\beta^3$ on the lattice.  Note
that it cannot be anymore a pointwise operator as for Wilson fermions
and staggered fermions -- it connects components which are located in
different points.  Thus, the original, continuous definition is not
enough to define $\gamma^5$ on the lattice.  But there is one very
natural choice: the shift operator in z-direction.  Modulo 2, on the
reference cube, it is indeed $\gamma^5$.  And, most important, it
anticommutes with the Dirac operator even on the lattice:

\begin{equation}
\gamma^5 D + D \gamma^5 = 0
\end{equation}

But, different from the $\gamma^5$ in continuous theory, it's square
is no longer 1, but a shift in z-direction.  This is a very nice and
beautiful way to break the exact chiral symmetry $(\gamma^5)^2=1$.
Note that $\gamma^5$ exchanges the even and odd fermion:

\begin{equation}
\gamma^5 \psi_e(n) = \psi_o(n-a_z)\hspace{1cm}
\gamma^5 \psi_o(n) = \psi_e(n-a_z)
\end{equation}

This gives the subspaces $\psi_\pm = \frac{1}{2}(1\pm\gamma^5)\psi$ a
quite obvious physical meaning: they are the subspaces defined by
$\psi_e\approx\pm\psi_o$.  On the other hand, because $\psi_e$ and
$\psi_o$ are located on different sub-meshs, we do not have a natural
symmetric and exact definition of the projectors.

Now, the most beautiful surprise we find thinking about the connection
to gauge fields.  As usual, we describe gauge degrees of freedom as
located on edges.  Now, a quite natural, ``naive'' symmetric lattice
interpolation for interaction terms $A\psi$ between fermions and the
gauge field is

\begin{equation}
A_{n+a_i,n}\frac{\psi(n+a_i)+\psi(n)}{2}
\end{equation}

Thus, we obtain a dependency on terms of type $\psi(n+a_i)+\psi(n)$ in
a natural way.  Now, these terms are $\approx \psi(n)$ for $\psi_+$
but $\approx 0$ for $\psi_-$ -- thus, our interaction terms with gauge
fields have exactly the $(1+\gamma^5)$ form of interaction we need in
chiral gauge theory!  Thus, in this discretization we have not only a
possibility to describe chiral gauge fields, but they appear almost by
themself, in a quite natural way.

This effect disappears if we have several components and the gauge
field connects them.  Thus, our $(1+\gamma^5)$ type interaction
appears only inside the doublet.  Outside the doublets we obtain a
vector-like interaction -- as it should be in chromodynamics.

\section{The Dirac Operator on the de Rham Complex}

The discretization may be generalized to the case of a general metric
background in a quite simple way -- as the exterior bundle or the de
Rham complex.  The de Rham complex and it's Dirac operator is
well-known, but has the ``wrong'' dimension for a single fermion: in
four dimension, we have $2^4 = 16$ components.  Instead, fermions are
usually described on a curved background using tetrad variables for
the metric (see, for example, \cite{Birrell}).

In our approach we have to describe a doublet of Dirac fermions, and
in this context these sixteen components of the exterior bundle is
exactly what we need.  In this sense, the ``fermion doubling'' problem
has a natural analogon in continuous theory on a curved background:
only a pair of fermions may be described on a curved metric background
$g_{\mu\nu}$ without tetrad variables.

\subsection{Hodge Theory}

Let's remember the basic formulas for the Dirac operator in the
exterior bundle (see, for example, \cite{Pete}).  The exterior bundle
or de Rham complex $\Omega = \sum_{k=0}^n \Omega^k$ consists
skew-symmetric tensor fields of type $(0,k), 0\le k \le n$ which are
usually written as differential forms

\begin{equation}
\psi = \psi_{i_1\ldots i_k} dx^{i_1}\wedge \cdots \wedge dx^{i_k} \in \Omega^k
\end{equation}

The exterior bundle $\Omega$ has dimension $2^n$ in the
n-dimensional space.  The most important operation on $\Omega$ is
the external derivative $d:\Omega^k\to\Omega^{k+1}$ defined by

\begin{equation}
(d\psi)_{i_1\ldots i_{k+1}}=\sum_{q=1}^{k+1}\frac{\partial}{\partial x^{i_q}}
 (-1)^q \psi_{i_1\ldots \hat{i}_q\ldots i_{k+1}} 
\end{equation}

where $\hat{i}_q$ denotes that the index $i_q$ has been omitted. It's
main property is $d^2=0$.  In the presence of a metric, we have also
the important $*$-operator $\Omega^k\to\Omega^{n-k}$:

\begin{equation}
(*\psi)_{i_{k+1}\ldots i_n} = \frac{1}{k!} \varepsilon_{i_1\ldots i_n} 
g^{i_1j_1} \cdots g^{i_kj_k}\psi_{j_1\ldots j_k}
\end{equation}

with $*^2 = (-1)^{k(n-k)}\mbox{sgn}(g)$.  This allows to define a
global inner product by

\begin{equation}
(\phi,\psi) = \int \phi \wedge (*\psi) = \int \psi \wedge (*\phi)
\end{equation}

It turns out that the adjoint operator of $d^*$ of $d$ is

\begin{equation}
d^* = (-1)^{rn+n+1} * d *
\end{equation}

In this general context we can define the Laplace operator as

\begin{equation}
\Delta = d d^* + d^* d
\end{equation}

Then, the Dirac operator (as it's square root) can be defined as

\begin{equation}
D = d+d^*.  
\end{equation}

Indeed, we have $d^2=0$ as well as $(d^*)^2=0$.  Sometimes the
${\mathbb Z}_2$ graduation is also useful: $\varepsilon \psi =
(-1)^k\psi$ if $\psi\in\Omega^k$.  The subspaces $\varepsilon=1$ and
$\varepsilon=-1$ have equal dimension $2^{n-1}$.
  
\subsection{Discretization on the de Rham Complex}

If we want to consider the approximation of some continuous object on
a general mesh, the exterior bundle is a very natural object. Indeed,
k-forms may be integrated over k-dimensional surfaces.  Thus, if a
general mesh (a cell complex) is given, a k-form defines a function on
the k-dimensional cells of the mesh in a very natural way. For each
cell we can define it's ``characteristic form'' on the mesh as the
equivalence class of forms with $\chi^i(c_j)=\delta^i_j$ and decompose
the general form on the mesh as $f(c)=f_i\chi^i(c)$.  The external
derivative defines in a similar natural way a derivative for functions
on the mesh, with the same most important exact property $d^2=0$.

For the definition of the $*$-operator we need a metric and a dual
mesh.  The metric defines in a natural way for every cell $c_i$ it's
area $a_i=a(c_i)> 0$.  In the Euclidean case and a triangulation,
these values depend on each other, but in the general case they may be
considered as independent variables which define the metric. In the
following we consider them as given and defining the metric.

The dual mesh is a quite natural mesh which can be defined for every
mesh: for each k-dimensional cell $c_i$ of the original mesh it has a
$(n-k)$-dimensional dual cell $\hat{c}_i$ which intersects the
original cell in a single point orthogonally and with positive
intersection index.  Now, the metric defines as well a scalar value
$\hat{a}_i$ for each cell of the dual mesh $\hat{c}_i$ -- the area of
the dual cell.  Now, the $*$-operator on the mesh may be defined as

\begin{equation}
(*f)(c)=\frac{\hat{a}_i}{a_i}f_i\chi^i(c)
\end{equation}

Note that the dual of the dual mesh has the same cells, but in some
dimensions with different orientations. Therefore, for $*^2$ there
appears the factor $(-1)^{k(n-k)}$ as in the continuous case.

Note also that the discrete exterior bundle is connected in a simple
way with a natural ``regular refinement'' step of a mesh: To refine a
mesh in a regular way, we can put one node in the center of each cell,
and connect nodes of neighbour dimension cells with edges if one of
the cell is on the boundary of the other.  This leads to a mesh where
each cell has the topological structure of a unit hypercube.

For a simple hypercubic lattice on a flat background, this
discretization scheme reduces to our proposal above.

\section{Conclusion}

The approach described here may help to solve some outstanding
problems of chiral gauge theory.  First, there is the general problem
how to formulate a chiral gauge theory on the lattice, known as the
``regularization problem'': any consistent regularization that
preserves the gauge symmetry must refer to the fermion
representation.\cite{Luescher}.  In our case, we describe chiral gauge
theory only in a very special, two-dimensional representation,
therefore it is not in conflict with this requirement.  And it gives a
representation we need in the standard model to describe the $SU(2)$
part of the electroweak interaction.  Moreover, if neutrinos are Dirac
particles, we don't need any other representations -- only their
combinations (one lepton and three quark doublets).  Of course, a lot
of things have to be done to obtain a complete quantum chiral gauge
theory in this way.  The problem how to handle the $U(1)$ part of
electroweak interaction, which needs the incorporation of all fermion
doublets, is completely open.

Second, it may be helpful for a solution of the
fermion doubling problem (see  \cite{Gupta} for an introduction,
[7,9,12,17] \cite{Kerler, Narayanan,Luescher,Testa} for various approaches to solve this problem
in chiral gauge theory).
It reduces the number of ``doubled flavours'' from four in the staggered
fermion approach (which seem unphysical) to two which already
allow a meaningful physical interpretation as lepton or quark
doublets. The nontrivial character of
the complex structure may prevent the application
of standard no-go-theorems like the famous Nielson-Ninomiya [10] theorem.

We also obtain an interesting representation of fermions on a general
metric background which does not need tetrad or triad variables for
the gravitational field.  Instead, the gravitational field may be
described as usual by the metric tensor $g_{\mu\nu}$.

\section{Acknowledgements}

Thanks to M. Nobes, L. Bourhis and S. Carlip for interesting
discussions.

\begin{appendix}

\section{Connection to the General Lorentz Ether}
\label{ether}

The approach to fermions presented here has been developed to solve a
particular problem connected with the description of fermion fields
General Lorentz Ether Theory (GLET), which we want to describe here
shortly.

This theory has been proposed by the author in \cite{Schmelzer} as a
generalization of the classical Lorentz ether to gravity.  It is a
metric theory of gravity, where the ``effective'' metric $g_{\mu\nu}$
is defined by a variant of the ADM decomposition from classical
condensed matter (ether) variables: density $\rho =
g^{00}\sqrt{-g}>0$, velocity $v^i=g^{0i}/g^{00}$, and a
three-dimensional pressure tensor $p^{ij}\sqrt{p} = -g^{ij}\sqrt{g}$.
The continuity and Euler equations of classical continuum mechanics

\begin{eqnarray}
\partial_t \rho + \partial_i (\rho v^i) &= &0 \label{continuity}\\
\partial_t (\rho v^j) + \partial_i(\rho v^i v^j + p^{ij}) &= &0 \label{Euler}.
\end{eqnarray}

transforms into the harmonic condition $\partial_\mu
g^{\mu\nu}\sqrt{-g}=0$ for the effective metric.  The Lagrangian of
GLET consists of the classical GR Lagrangian and some non-covariant
term which describes the dependence on the Newtonian background:

\begin{equation}\label{L}
L = L_{GR} + L_{matter}(g_{\mu\nu},\varphi^m)
  - (8\pi G)^{-1} (\Upsilon g^{00}-\Xi \delta_{ij}g^{ij})\sqrt{-g},
\end{equation}

Therefore, the theory has a well-defined GR limit $\Upsilon,\Xi\to 0$
which makes the theory viable.  This Lagrangian has been derived from
first principles which seem quite natural for condensed matter
theories in \cite{Schmelzer}.  Especially the covariant character of
its matter part, that means the Einstein equivalence principle,
follows.

It has been noted \cite{Bourhis} that this derivation does not extend
(at least not in any obvious way) to tetrad or triad variables.
Because the standard way to describe fermions on a curved background
uses tetrad (or, in the context of an ADM decomposition, triad)
variables, this makes the incorporation of fermions into this theory
problematic.  One way to solve this problem is to find a way to
describe fermions without tetrad or triad variables, what has been
done in our approach: Here we define a pair of Dirac fermions as 
classical tensor fields.

It should also be noted that the regularization problem of chiral
gauge theory as well as the fermion doubling problem become important
in ether theory if we want to use an ``atomic ether'' concept to
regularize the theory.  An ``atomic ether'' regularization, which
interprets the density $\rho$ as a particle density and velocity $v^i$
as their average velocity is a quite natural way to quantize an ether
theory.  But in this case we need some lattice discretization for all
parts of the standard model \cite{Nobes}.  

In the context of the ether approach, an extremely simple toy model
would be a cubic ether crystal.  Distortions of this crystal would be
defined by three real values per node.  This toy model exactly fits
into the lattice discretization described here, which gives three
generations of a pair of fermions.  Thus, already an extremely simple
toy model seems to be able to explain the three fermion generations.
Moreover it suggests that fermions appear in pairs, with a natural
possibility for $1+\gamma^5$ type gauge interactions inside these
pairs.  These are already nontrivial and important features of the
standard model.  Especially if neutrinos appear to be Dirac particles,
this scheme holds for all fermions of the standard model.

\section{Compatibility with ADM Decomposition}

For the compatibility of the approach described here with GLET it is
the compatibility with the ADM decomposition is crucial and worth to
be considered.  Using ``comoving'' spatial coordinates (which remain
constant along the velocity field $v^i$), the harmonic operator of the
metric $g_{\mu\nu}$ reduces to

\begin{equation}
\square \psi = -(\rho \partial_t^2 - \Delta) \psi
\end{equation}

where $\Delta$ is the standard three-dimensional (harmonic) Laplace
operator of the metric $p_{ij}$.  Thus, we ADM decomposition allows to
reduce the harmonic equation to a ``three-and-one-half-dimensional''
equation

\begin{equation}
\sqrt{\rho}\partial_t\psi = \sqrt{\Delta} \psi.
\end{equation}

Now it seems natural to use a mesh which is compatible with the ADM
decomposition.  We start with an arbitrary mesh on the spatial slice
at a given time $t_0$.  On this three-dimensional mesh we can define
the Dirac operator following the scheme described above.  Then we can
extend this mesh into a four-dimensional mesh.  First, we extend the
mesh nodes at $t_0$ following the comoving spatial coordinates into
comoving world-lines.  For the preferred time we use a global, regular
lattice and simple central differences.  Such a choice of a lattice
and discretization is in natural agreement with a crystallic ether
model.

\section{A Natural $SO(3)$ Gauge Field for a General
Three-Dimensional Mesh}

Another reason why we think that the focus of three spatial dimensions
will be helpful is that a general three-dimensional mesh defines in a
simple and natural way an $SO(3)$ gauge field.  Because at the Lie
algebra level $so(3)=su(2)$ this may be a candidate for the $su(2)$
gauge field of the electroweak interaction.

To understand this gauge connection, remember that to define a gauge
connection it is sufficient to define its integral over closed loops.
Thus, let's define the integral over a closed loop starting from a
general point (that means, inside a cell) and of general type (thus,
the loop intersects only planar cells of the mesh).  

Now, assume also that we have made a simple regular refinement step,
therefore, the mesh is a cubic mesh.  For each cubic cell, we fix some
standard map into the regular cube.  Then we start with a standard
reper (as it comes from the standard cube) and use trivial parallel
transport of the reper (as defined on the standard cube) inside the
cells up to the boundary.  On the boundary, one vector of the reper is
always orthogonal to the boundary.  If we come back to the original
node, the transferred reper is also, by construction, a reper in this
cube, its directions are parallel to the original reper.  Nonetheless,
it is in general a rotated reper.  This rotation defines the gauge
connection.  It measures the irregularity of the mesh: For a regular
cubic mesh it is trivial.

Now, the existence of a gauge field which is in such a natural way
connected with mesh irregularity does not look like an accident of
nature.  Our hypothesis is that it is (at least connected in some way
with) the $SU(2)$ gauge field of the electroweak interaction.  The
details of such a connection are subject of future research.

\end{appendix}

\end{document}